\documentclass[english,onecolumn,superscriptaddress,notitlepage,nofootinbib]{revtex4-1}
\usepackage[T1]{fontenc}
\usepackage[latin9]{inputenc}
\usepackage{color}
\usepackage{babel}
\usepackage{amsmath}
\usepackage{graphicx}
\usepackage[unicode=true,pdfusetitle,
 bookmarks=true,bookmarksnumbered=false,bookmarksopen=false,
 breaklinks=false,pdfborder={0 0 0},pdfborderstyle={},backref=false,colorlinks=true]
 {hyperref}
\hypersetup{
 linkcolor=coolblack,citecolor=burntorange,urlcolor=burntgreen}

\makeatletter

\newcommand{\lyxdot}{.}

\usepackage{color}
\definecolor{burntorange}{rgb}{0.8, 0.33, 0.0}
\definecolor{charcoal}{rgb}{0.21, 0.27, 0.31}
\definecolor{coolblack}{rgb}{0.0, 0.28, 0.49}
\definecolor{burntgreen}{rgb}{0.05, 0.45, 0.27}

\makeatother

\begin{document}

\title{A method to reduce the rejection rate in Monte Carlo Markov Chains}

\author{Carlo Baldassi}

\affiliation{Department of Applied Science and Technology, Politecnico di Torino,
Corso Duca degli Abruzzi 24, I-10129 Torino, Italy}

\affiliation{Human Genetics Foundation-Torino,  Via Nizza 52, I-10126 Torino,
Italy}
\begin{abstract}
We present a method for Monte Carlo sampling on systems with discrete
variables (focusing in the Ising case), introducing a prior on the
candidate moves in a Metropolis-Hastings scheme which can significantly
reduce the rejection rate, called the reduced-rejection-rate (RRR)
method. The method employs same probability distribution for the choice
of the moves as rejection-free schemes such as the method proposed
by Bortz, Kalos and Lebowitz (BKL)~\cite{bortz_new_1975}; however,
it uses it as a prior in an otherwise standard Metropolis scheme:
it is thus not fully rejection-free, but in a wide range of scenarios
it is nearly so. This allows to extend the method to cases for which
rejection-free schemes become inefficient, in particular when the
graph connectivity is not sparse, but the energy can nevertheless
be expressed as a sum of two components, one of which is computed
on a sparse graph and dominates the measure. As examples of such instances,
we demonstrate that the method yields excellent results when performing
Monte Carlo simulations of quantum spin models in presence of a transverse
field in the Suzuki-Trotter formalism, and when exploring the so-called
robust ensemble which was recently introduced in~\cite{baldassi2016_unreasonable}.
Our code for the Ising case is publicly available~\cite{RRR-code},
and extensible to user-defined models: it provides efficient implementations
of standard Metropolis, the RRR method, the BKL method (extended to
the case of continuous energy specra), and the waiting time method~\cite{dall_faster_2001}.

\tableofcontents{}
\end{abstract}
\maketitle

\section{Introduction}

Monte Carlo methods play a central role in the simulation and study
of a variety of physical and mathematical problems. In particular,
Monte Carlo Markov Chains (MCMC) are a class of very powerful and
general algorithms to sample from arbitrary probability distributions~\cite{krauth_statistical_2006};
furthermore, they are at the basis of some general optimization techniques
such as simulated annealing~\cite{kirkpatrick1983optimization} and
quantum annealing~\cite{santoro2002theory}. Arguably, the most popular
general MCMC scheme is based on the Metropolis-Hastings rule. Concrete
implementations of the rule can however suffer, particularly in frustrated
systems, from the problem of having a very slow dynamics; one typical
manifestation of this is a very high rejection rate of the proposed
moves. Indeed, a wealth of different techniques have been proposed
in order to overcome this problem and improve the sampling efficiency
for specific classes of systems, such as Kinetic Monte Carlo~\cite{voter2007introduction},
Cluster Monte Carlo~\cite{wang1990cluster,wolff1989collective},
parallel tempering \cite{earl2005parallel}, and many others (see
e.g.~\cite{landau_guide_2009}).

In this paper, we propose yet another such variant, which we call
``reduced-rejection-rate Monte Carlo'', RRR for short, with a publicly
available generic implementation~\cite{RRR-code}, and present some
exploratory numerical results that demonstrate its advantages. The
method as described here is applicable to the case of Ising spin systems
but could be generalized to Potts-like models, and consists in introducing
a prior on the choice of the candidate moves in an otherwise standard
Metropolis-Hastings scheme, with the aim \textendash{} as the name
suggests \textendash{} of reducing the rejection rate.

Consider a system of $N$ interacting Ising spins $\sigma=\left\{ \sigma_{i}\right\} _{i=1}^{N}\in\left\{ -1,+1\right\} ^{N}$
subject to some Hamiltonian $E\left(\sigma\right)$. For simplicity,
in the following we will consider Hamiltonians restricted to at most
pairwise interactions between the spins, but the treatment is general.
Let us then call $\mathcal{L}$ the set of interacting pairs, and
write:
\begin{equation}
E\left(\sigma\right)=-\sum_{\left(i,j\right)\in\mathcal{L}}J_{ij}\sigma_{i}\sigma_{j}+\sum_{i}h_{i}\sigma_{i}.\label{eq:E}
\end{equation}

We want to sample from the Boltzmann probability distribution at inverse
temperature $\beta$:
\begin{equation}
P\left(\sigma\right)=\frac{e^{-\beta E\left(\sigma\right)}}{Z}.
\end{equation}

In the usual MCMC scheme, using the Metropolis-Hastings rule, one
starts from a configuration $\sigma$, proposes a candidate move $\sigma\to\sigma^{\prime}$
with some prior probability distribution $C\left(\sigma\to\sigma^{\prime}\right)$,
and accepts the move with some acceptance rate $A\left(\sigma\to\sigma^{\prime}\right)$.
Therefore, we write the transition probability of going from configuration
$\sigma$ to a different configuration $\sigma^{\prime}$ in the Markov
Chain as:
\begin{equation}
P\left(\sigma\to\sigma^{\prime}\right)=C\left(\sigma\to\sigma^{\prime}\right)A\left(\sigma\to\sigma^{\prime}\right)\qquad\sigma^{\prime}\ne\sigma.
\end{equation}
Of course, $P\left(\sigma\to\sigma\right)=1-\sum_{\sigma^{\prime}\ne\sigma}P\left(\sigma\to\sigma^{\prime}\right)$.

The rejection rate is determined by enforcing the detailed balance
condition $P\left(\sigma\right)P\left(\sigma\to\sigma^{\prime}\right)=P\left(\sigma^{\prime}\right)P\left(\sigma^{\prime}\to\sigma\right)$:
\begin{equation}
\frac{A\left(\sigma\to\sigma^{\prime}\right)}{A\left(\sigma^{\prime}\to\sigma\right)}=e^{-\beta\left(E\left(\sigma^{\prime}\right)-E\left(\sigma\right)\right)}\frac{C\left(\sigma^{\prime}\to\sigma\right)}{C\left(\sigma\to\sigma^{\prime}\right)}
\end{equation}
which can be accomplished by the usual formula:
\begin{equation}
A\left(\sigma\to\sigma^{\prime}\right)=\min\left(1,e^{-\beta\left(E\left(\sigma^{\prime}\right)-E\left(\sigma\right)\right)}\frac{C\left(\sigma^{\prime}\to\sigma\right)}{C\left(\sigma\to\sigma^{\prime}\right)}\right).\label{eq:rejection}
\end{equation}

In a straightforward implementation, the proposed moves consist in
choosing one spin uniformly at random and flipping it. Then, the $C$
terms in the above equation simplify and one is left with the simple
rule:
\begin{equation}
P\left(\sigma\to\sigma^{\prime}\right)=\min\left(1,e^{-\beta\left(E\left(\sigma^{\prime}\right)-E\left(\sigma\right)\right)}\right).\label{eq:standard}
\end{equation}

Throughout the paper, we refer to this choice as a ``standard Metropolis
scheme''. In this work, we will instead focus on choosing the prior
$C$ in such a way to reduce the rejection rate, while still keeping
its computation efficient. The rejection rate minimization is achieved,
informally speaking, by making the quantity 
\begin{eqnarray}
R\left(\sigma\to\sigma^{\prime}\right) & = & e^{-\beta\left(E\left(\sigma^{\prime}\right)-E\left(\sigma\right)\right)}\frac{C\left(\sigma^{\prime}\to\sigma\right)}{C\left(\sigma\to\sigma^{\prime}\right)}\label{eq:R}
\end{eqnarray}
 as close as possible to $1$. We will still only consider single
flips as the candidate moves.

The basis of our method is (a generalization of) the rejection-free
algorithm by Bortz, Kalos and Lebowitz~\cite{bortz_new_1975}, which
we refer to as BKL throughout this paper (it is also known as ``the
n-fold way''). The BKL method goes under the general category of
``Kinetic Monte Carlo'', or ``faster-than-the-clock'' (FTTC) schemes,
see e.g.~\cite{krauth_statistical_2006}. The general FTTC technique
is to pre-compute the probability distribution of the number of consecutive
rejections before a move would be accepted, extract a number of MCMC
iterations to skip from that distribution (thus ``advancing the clock''),
then choose a move without rejections. The resulting MCMC has the
same statistics as a standard Metropolis scheme, but skipping the
rejections may be computationally convenient, especially at low temperatures
when the rejection rate is high. Of course, the additional computations
involved only lead to an advantage in some cases. In particular, the
BKL method was introduced as an efficient method for the case of Ising
spin systems with small connectivity, and in which the energy shift
$\Delta E\left(\sigma\to\sigma^{\prime}\right)=E\left(\sigma^{\prime}\right)-E\left(\sigma\right)$
induced by a spin flip can only belong to a small set of values. As
an example of such system, consider an Edwards-Anderson model, i.e.~a
$D$-dimensional lattice with nearest neighbor interactions and periodic
boundary conditions, with $J_{ij}\in\left\{ -1,+1\right\} $ and $h_{i}=0$,
in which case there are only $2D+1$ possible values of $\Delta E$:
$\Delta E\in\left\{ -4D,-4D+4,\dots,4D-4,4D\right\} $. As it turns
out, the requirement of small connectivity $K\ll N$ is the only crucial
one in order to achieve an efficient implementation, while the requirement
that the energy shifts are discrete is useful for further optimizing
the method: for general energy shifts, an efficient rejection-free
method exists, the waiting time method (WTM for short)~\cite{dall_faster_2001},
but we will show that the BKL can also be extended rather straightforwardly
and achieve comparable performances to WTM. We will shortly review
the BKL method in section~\ref{sec:BKL}, and show how to modify
it in the case of general heterogeneous energy shifts.

The core of our RRR method is to use the same probability distribution
for the choice of the move as in the BKL method, but this time as
the prior $C$, without the initial skipping of the rejected moves;
this leaves a residual rejection rate, given by eq.~(\ref{eq:rejection}).
Although this may appear to be a net loss, this is often not the case,
and may even be slightly advantageous in a number of situations, as
shown in section~\ref{subsec:Tests-BKL}. The main advantage of the
RRR method however is that it is more easily generalizable. Consider
the case in which the energy can be written as the sum of two components:
\begin{equation}
E\left(\sigma\right)=E_{s}\left(\sigma\right)+E_{d}\left(\sigma\right)
\end{equation}
where $E_{s}\left(\sigma\right)$ describes a model with low connectivity,
while $E_{d}\left(\sigma\right)$ is a residual part of the energy
for which the BKL or WTM methods are not convenient. With our approach,
since \textendash{} as we shall show \textendash{} we can achieve
\begin{equation}
e^{-\beta\left(E_{s}\left(\sigma^{\prime}\right)-E_{s}\left(\sigma\right)\right)}\frac{C\left(\sigma^{\prime}\to\sigma\right)}{C\left(\sigma\to\sigma^{\prime}\right)}\approx1
\end{equation}
we are then only left with the acceptance rate relative to the non-sparse
part of the system:
\begin{equation}
A\left(\sigma\to\sigma^{\prime}\right)\approx\min\left(1,e^{-\beta\left(E_{d}\left(\sigma^{\prime}\right)-E_{d}\left(\sigma\right)\right)}\right).
\end{equation}

Then, with the RRR method, we can almost completely absorb the effect
of the interaction $E_{s}$ by including it in the prior. This basic
idea was used, with the method we propose here in embryonic form,
when studying the so-called robust ensemble (RE) introduced in ref.~\cite{baldassi2016_unreasonable};
the RRR method is a generalization and an improvement of what was
used in that paper. This case is discussed in section~\ref{subsec:Tests-RE}.
An important case that also falls in this category is the simulation
of quantum spin systems via the Suzuki-Trotter transformation~\cite{suzuki1976relationship},
in which case $E_{s}$ can be identified with the interactions between
the replicated Suzuki-Trotter spins. This case is discussed in section~\ref{subsec:Tests-Quantum},
where indeed we will show that RRR is able to equilibrate much faster
than the standard Metropolis scheme.

\section{The BKL Method\label{sec:BKL}}

In this section, we briefly review the BKL method. In its original
formulation, the authors called the method ``the n-fold way'', since
they were considering the case in which the energy shifts arising
from a spin flip could only take a value in a small discrete set.
Here, however, we will start with the more general case, and recover
the original formulation as a specialized case. We will thus introduce
some definitions and notations which we will used for RRR as well.

Let us indicate by $\sigma^{\left(i\right)}$ a configuration of the
spins obtained from another configuration $\sigma$ by flipping the
spin $i$:
\begin{equation}
\sigma_{j}^{\left(i\right)}=\begin{cases}
\sigma_{j} & \mathrm{if}\,j\ne i\\
-\sigma_{j} & \mathrm{if}\,j=i.
\end{cases}
\end{equation}

We denote the effect of such a spin flip on the energy as:
\begin{equation}
\Delta E_{\sigma}^{\left(i\right)}=E\left(\sigma^{\left(i\right)}\right)-E\left(\sigma\right).
\end{equation}

Then, for a given configuration $\sigma$ and all possible spin flips,
we define the following quantities:
\begin{eqnarray}
p_{\sigma}^{\left(i\right)} & = & \min\left(1,e^{-\beta\Delta E_{\sigma}^{\left(i\right)}}\right)\label{eq:p}\\
z_{\sigma} & = & \sum_{i}p_{\sigma}^{\left(i\right)}.\label{eq:z}
\end{eqnarray}

With these, the probability of rejecting a move in a standard Metropolis
scheme (eq.~(\ref{eq:standard})) is $P\left(\sigma\to\sigma\right)=1-z_{\sigma}/N$,
while $P\left(\sigma\to\sigma^{\left(i\right)}\right)=p_{\sigma}^{\left(i\right)}/N$
is the probability of transitioning to the new configuration $\sigma^{\left(i\right)}$.
The BKL procedure at each step is then as follows: 
\begin{enumerate}
\item Extract a number of iterations to skip as $\left\lfloor \frac{\mbox{\ensuremath{\log}}\left(1-r\right)}{\log\left(1-z_{\sigma}/N\right)}\right\rfloor $,
where $r$ is a random number extracted uniformly in $\left[0,1\right)$,
and ``advance the clock'' accordingly.
\item Extract a spin $i$ with probability $p_{\sigma}^{\left(i\right)}/z_{\sigma}$
and flip it, thereby changing the configuration to $\sigma^{\left(i\right)}$.
\item Compute the new values $p_{\sigma^{\left(i\right)}}^{\left(j\right)}$
for all $j$ and the new $z_{\sigma^{\left(i\right)}}$.
\end{enumerate}
As mentioned above, by this procedure one realizes the same MCMC transition
matrix of standard Metropolis, but can save computational time by
skipping the rejected iterations entirely (step 1), at the cost of
requiring a specialized sampling procedure (step 2) and some additional
bookkeeping (step 3). Note that in step 3 only the flipped spin and
its neighbors may change their values of $p_{\sigma}^{\left(i\right)}$,
therefore the bookkeeping operations can be relatively inexpensive
for diluted graphs (the details on how to achieve this are given in
section~\ref{sec:Implementation}), which explains the reason for
the requirement $K\ll N$ given in the introduction.

When expressed in this general form, the BKL method is almost equivalent
to another closely related rejection-free method, the waiting time
method (WTM)~\cite{dall_faster_2001}. However, in the special case
in which the set $\left\{ \Delta E_{\sigma}^{\left(i\right)}\right\} $
has a small cardinality for all $\sigma$ and all $i$ (e.g. in the
case of a regular lattice with $\pm1$ couplings mentioned in the
introduction), a further optimization can provide additional computational
advantages. In that case, which is the one originally considered by
the authors in~\cite{bortz_new_1975}, the procedure is as follows:
for a given configuration $\sigma$, we divide all the spins into
classes, based on the energy shift induced by flipping them:
\begin{equation}
\mathcal{C}_{\sigma}\left(\Delta E\right)=\left\{ i:E\left(\sigma^{\left(i\right)}\right)-E\left(\sigma\right)=\Delta E\right\} .
\end{equation}

All the spins in the same class will thus have the same associated
probability $p_{\sigma}^{\left(i\right)}$ (eq.~(\ref{eq:p})), thus
we only need to keep track of one probability for each class, and
of the class sizes. Let us then define:
\begin{eqnarray}
n_{\sigma}\left(\Delta E\right) & = & \left|\mathcal{C}_{\sigma}\left(\Delta E\right)\right|\label{eq:n}\\
p_{\sigma}\left(\Delta E\right) & = & n_{\sigma}\left(\Delta E\right)\min\left(1,e^{-\beta\Delta E}\right).\label{eq:p_deltaE}
\end{eqnarray}

With these, we can also express $z_{\sigma}$ (eq.~(\ref{eq:z}))
as: 
\begin{equation}
z_{\sigma}=\sum_{\Delta E}p_{\sigma}\left(\Delta E\right).\label{eq:z_deltaE}
\end{equation}

In this specialized case, the step 2 of the BKL scheme above is performed
in two separate steps:
\begin{enumerate}
\item [2.]\setcounter{enumi}{2}
\begin{enumerate}
\item Extract a class $\mathcal{C}_{\sigma}\left(\Delta E\right)$ with
probability $p_{\sigma}\left(\Delta E\right)/z_{\sigma}$
\item Extract a spin uniformly at random from the chosen class, and flip
it.
\end{enumerate}
\end{enumerate}
The update step 3 is also simplified, since it can use a more efficient
data structure. The details are given in section~\ref{sec:Implementation}.

\section{The RRR Method\label{sec:RRR}}

In the RRR method, when applied to sparse models, the proposal $C\left(\sigma\to\sigma^{\prime}\right)$
simply follows the BKL step 2 of the previous section, without the
skipping step 1 and without accepting the move right away, i.e.:
\begin{equation}
C\left(\sigma\to\sigma^{\left(i\right)}\right)=\frac{p_{\sigma}^{\left(i\right)}}{z_{\sigma}}.
\end{equation}

With this choice, the expression involved in the acceptance rate $A\left(\sigma\to\sigma^{\left(i\right)}\right)$,
eq.~(\ref{eq:R}), takes the remarkably simple form:
\begin{eqnarray}
R\left(\sigma\to\sigma^{\left(i\right)}\right) & = & \frac{z_{\sigma}}{z_{\sigma^{\left(i\right)}}}\label{eq:rejection_explicit}
\end{eqnarray}
where $z_{\sigma}$ is the same as for the BKL algorithm, eq.~(\ref{eq:z}),
and $z_{\sigma^{\left(i\right)}}$ is the same quantity computed for
the new candidate configuration $\sigma^{\left(i\right)}$. Indeed,
this choice significantly reduces \textendash{} and in many cases
nearly eliminates \textendash{} the rejection rate. The basic reason
for this is that, at fixed (non-zero) temperature and in the thermodynamic
limit $N\to\infty$, $z_{\sigma}$ is an extensive quantity and the
perturbation induced by the spin flip is at most of the order of the
connectivity $K$, therefore $z_{\sigma^{\left(i\right)}}=z_{\sigma}+\mathcal{O}\left(K\right)$
and $R\left(\sigma\to\sigma^{\left(i\right)}\right)\approx1$ to the
leading order. Only at fixed $N$ and very low temperatures the difference
becomes more significant (and indeed, in the limit of $\beta\to\infty$
at fixed $N$ the rejection rate must tend to $1$ whenever the system
is in a local minimum); in practice, though, this regime appears to
be rather narrow, while the acceptance rate seems to be very close
to $1$ up to rather high values of $\beta$ (see the numerical experiments
of section~\ref{subsec:Tests-BKL}). Moreover, consider a system
initialized in a random configuration in the initial transient regime,
at an energy $E\left(\sigma\right)$ far above the equilibrium one:\footnote{And also far above that of any long-lived meta-stable state which
may trap the dynamics, should they exists.} the proposed move $i$ will be more likely associated to a negative
energy shift $\Delta E_{\sigma}^{\left(i\right)}<0$, due to the form
of $p_{\sigma}^{\left(i\right)}$ in eq.~(\ref{eq:p}). In that case,
the contribution of spin $i$ to the normalization term $z$ is $1$
in $z_{\sigma}$ and it is $e^{\beta\Delta E_{\sigma}^{\left(i\right)}}$
in $z_{\sigma^{\left(i\right)}}$, and thus $z_{\sigma^{\left(i\right)}}<z_{\sigma}$
unless the effect is overcome by the shifts on the neighbors, which
means that in the initial stages there is a bias towards $R\left(\sigma\to\sigma^{\left(i\right)}\right)>1$,
further reducing the rejection rate.

The resulting MCMC transition matrix is thus no longer the same as
that of standard Metropolis, and has in general better convergence
properties (for very small systems, this can be assessed numerically
by computing the eigenvalues of the transition matrices).

When the method is applied to the two-level systems mentioned in the
introduction, i.e.~systems such that the energy can be written as
$E\left(\sigma\right)=E_{s}\left(\sigma\right)+E_{d}\left(\sigma\right)$
with only $E_{s}\left(\sigma\right)$ begin sparse, the RRR algorithm
is straightforwardly modified as follows: first, a candidate spin
$i$ is chosen according to $E_{s}\left(\sigma\right)$ as described
above, and $R\left(\sigma\to\sigma^{\left(i\right)}\right)$ is computed
from eq.~(\ref{eq:rejection_explicit}); then, the residual energy
shift $\Delta E_{d\sigma}^{\left(i\right)}=E_{d}\left(\sigma^{\left(i\right)}\right)-E_{d}\left(\sigma\right)$
is computed; finally the move is accepted with probability
\begin{equation}
A\left(\sigma\to\sigma^{\left(i\right)}\right)=\min\left(1,R\left(\sigma\to\sigma^{\left(i\right)}\right)e^{-\beta\,\Delta E_{d\sigma}^{\left(i\right)}}\right).
\end{equation}

When $\Delta E_{s}$ is typically much larger than $\Delta E_{d}$,
this method can provide very significant improvements. Experiments
on this type of systems are shown in sections~\ref{subsec:Tests-Quantum}
and~\ref{subsec:Tests-RE}.

In principle, it may also be possible to exploit the fact that discrete
energy spectra allow for a specialized version of the algorithm, and
apply the RRR method to sparse graphs with continuous spectra by discretizing
the interactions and allowing for a small residual rejection rate
(e.g.~by writing the couplings as $J_{ij}=J_{ij}^{0}+\delta J_{ij}$
with the $J_{ij}^{0}$ chosen from a small discrete set and such that
$\left|\delta J_{ij}\right|$ is small). In our tests, however, this
strategy led at best to the same performances as using the more general
versions of BKL/RRR, or WTM, i.e. the slight advantage of the discretization
was always at least counterbalanced by the disadvantage of having
a slightly increased rejection rate, and we have thus abandoned this
line of inquiry.

\section{Implementation details\label{sec:Implementation}}

During each step of the BKL or RRR algorithms, we need to have an
efficient way to \emph{a)} sample from the distribution $p_{\sigma}^{\left(i\right)}/z_{\sigma}$,
and \emph{b)} update the distribution under the assumption that only
a small subset $K\ll N$ of the $p_{\sigma}^{\left(i\right)}$ will
have changed. To achieve this in our implementation, we used a specialized
binary tree which keeps ``local cumulative distributions'' in $\mathcal{O}\left(N\right)$
memory and can be accessed and updated in $\mathcal{O}\left(\log N\right)$
time~\cite{wong_efficient_1980}. More precisely, let us at first
assume for simplicity that $N$ is a power of $2$, and define the
table of arrays $\left\{ a_{l}\right\} $, with $l=1,\dots,\log_{2}N$,
of variable length $2^{l-1}$; in each entry of an array $a_{l}$,
we store the sum of $N/2^{l}$ elements:
\begin{eqnarray}
a_{l}\left(k\right) & = & \sum_{j=1}^{N/2^{l}}p_{\sigma}^{\left((k-1)N/2^{l-1}+j\right)}\qquad k=1,\dots,2^{l-1}.
\end{eqnarray}
We also compute and store $z_{\sigma}=\sum_{i=1}^{N}p_{\sigma}^{\left(i\right)}$.
To extract a random element $i$ with probability $p_{\sigma}^{\left(i\right)}/z_{\sigma}$,
we use this procedure:
\begin{enumerate}
\item extract a random number $x\in\left[0,z_{\sigma}\right)$, and initialize
$k\leftarrow0$;
\item looping over each $l=1,\dots,\log_{2}N$, do:
\begin{enumerate}
\item if $x>a_{l}\left(k+1\right)$, set $k\leftarrow2k+1$ and $x\leftarrow x-a_{l}\left(k+1\right)$
\item otherwise set $k\leftarrow2k$;
\end{enumerate}
\item return $i=k+1$.
\end{enumerate}
This requires $\mathcal{O}\left(\log N\right)$ elementary operations.
Updating the structure when one of the $p_{\sigma}^{\left(i\right)}$
is changed is also $\mathcal{O}\left(\log N\right)$ since each entry
only appears in at most $\log_{2}N$ entries in the $\left\{ a_{l}\right\} $
table, and $z_{\sigma}$ can be updated with a single operation.\footnote{Other schemes based on more sophisticated data structures such as~\cite{matias_dynamic_2003}
can achieve better asymptotic performances, but they involve more
complex operations, and also consume more random numbers. We consider
it unlikely that they could positively and significantly affect the
results which we present here considering the system sizes involved
(and in any case this would only affect the comparisons between BKL/RRR
and WTM for the continuous cases), and we have thus left their use
as a potential future improvement to the code.} In the general case in which $N$ is not a power of $2$, we simply
pad the distribution with zeros. Also note that the WTM has the same
complexity, since it uses a binary heap as its underlying data structure
for performing the analogous sampling and update operations.

In the specialized case of discrete energy shifts, however, we use
basically the same approach and data structures as described in the
original BKL paper~\cite{bortz_new_1975}: at each time, we keep
track of the $\mathcal{C}\left(\Delta E\right)$ classes' compositions
by using unsorted lists of variable size, and associated look-up tables
to determine the position of each spin in the structure. Updating
the position of a spin within the structure is an $\mathcal{O}\left(1\right)$
operation, since the structure is unsorted. For example, removing
a spin from a list amounts at doing the following: locate its position
using the look-up table, move the last spin of the list in its position,
reduce the list size, update the look-up table. This data structure
then allows to \emph{a)} keep track of the values of $n_{\sigma}$
(eq.~(\ref{eq:n})) for all classes, \emph{b)} have an efficient
way to choose a spin within a class, and \emph{c)} determine how the
neighbors' classes are affected by the move and perform the update.
Note that when the number of classes is very small, the choice of
the class (step 2a at the end of section~\ref{sec:BKL}) is most
efficiently performed by simply extracting a uniform random number
in $\left[0,1\right)$ and computing the cumulative distribution on
the fly until it exceeds that number, without the need of specialized
structures. This is for cache efficiency reasons. When instead the
number of classes is not so small that more sophisticated methods
are required, the discrete specialization is not particularly convenient
over the more general method.

When determining the effect of the move on the energy shifts, a loop
over the neighbors of the chosen spin is necessary. This introduces
an $\mathcal{O}\left(K\right)$ cost, where $K$ is the connectivity,
as mentioned above. The difference between BKL and RRR is that, in
the latter case, this computation is required before the move is accepted.
For RRR, we have thus two possible modes of operation: we can either
always accept the change at first and undo it if it happens to be
rejected (we call this the \emph{undo mode}), or we can keep in memory
the results of the computation (the potential shifts induced by the
move) and apply them later if the move is accepted (we call this the
\emph{staged mode}). Empirically, we found that the undo mode has
better performance at high acceptance rate, while the staged mode
becomes convenient at lower acceptance rates (the transition between
the two regimes happening at about $0.5$ or $0.8$ acceptance rate
on the discrete or continuous problems we tested, respectively). Therefore,
our code determines which strategy to employ based on an ongoing estimate
of the acceptance rate. When the acceptance rate of RRR is high, these
additional computations are very rarely wasted with respect to the
BKL method \textendash{} it is rare that a move needs to be undone;
on the other hand, the RRR method avoids the computation of the number
of steps to skip (point 1 in section~\ref{sec:BKL}), which, as it
turns out, can also be relatively expensive. On balance, RRR may be
slightly more convenient than BKL even on diluted models, up to some
value of $\beta$ where the acceptance rate of RRR starts decreasing.

Finally, we mention the fact that it is common practice in Monte Carlo
simulations to boost performance by keeping some kind of memory cache
 (e.g. the local fields acting on a spin); the need, at least in the
staged mode of operation, to perform tentative spin flips, then roll
them back, and finally perhaps accepting them, introduces a slight
complication in this regard. This however is solved quite easily by
keeping an additional cache level which allows to undo the last performed
change: in our tests, this proved to be both computationally cheap
and sufficient to keep the advantages of the cache mechanisms. We
refer the interested reader to the provided code~\cite{RRR-code}
for concrete examples.

\section{Numerical experiments\label{sec:Numerical}}

We performed numerical experiments comparing the performance of standard
Metropolis, BLK (where applicable), WTM (where applicable) and RRR.
All tests were run under identical conditions on a 2.5 GHz Intel i7-4710HQ
CPU with a single thread; the code is written in the Julia programming
language and run with Julia version 0.5 on a Linux operating system.\footnote{For reference, we benchmarked the performance of our generic Julia
code against a very efficient specialized C implementation of BKL
on a p-spin model which was kindly made available to us by G.~Parisi,
and found them to be identical.}

In all tests, we set a hard wall-clock-time limit (e.g.~$60$ seconds),
and compared most results (e.g.~the energy or the overlaps) as a
function of the wall-clock time. The results were sampled at regular
intervals in the simulated Monte Carlo time: we took one sample for
every $10^{4}$ attempted moves in the RRR case, and we scaled the
sampling intervals for the other algorithms in order to make sure
that a similar amount of samples was collected for each algorithm.
We mostly explored the low-temperature regime, which is the one where
RRR (as well as BKL and WTM) can be expected to offer an advantage
over standard Metropolis.

\subsection{Tests on sparse models\label{subsec:Tests-BKL}}

The first batch of experiments which we present here was performed
on random regular graphs (RRG) with connectivity $K=3$, with random
couplings. We tested two cases, binary couplings $J_{ij}\in\left\{ -1,+1\right\} $
and continuous normally distributed couplings $J_{ij}\sim\mathcal{N}\left(0,1\right)$,
extracted uniformly and independently, in both cases with no external
fields ($h_{i}=0$, see eq.~(\ref{eq:E})).\footnote{Qualitatively similar results as those shown here were obtained with
Edwards-Anderson graphs and p-spin models with random couplings.} For each case, we tested at least $20$ different random graphs,
and ran $2$ independent tests for each algorithm and each setting
of the parameters. We performed all experiments at fixed $\beta$,
starting from random initial conditions. By construction, we were
not expecting the RRR method to outperform BKL or WTM on this type
of graph. Thus, the main purpose of these experiments was to check
the behavior of RRR, in particular as compared to BKL and WTM, exploring
different values of $\beta$ and $N$. As expected, at large $N$
and not-too-large $\beta$, RRR has a very high acceptance rate (e.g.~larger
than $0.99$ for $\beta=2$ in both models with $N=10^{4}$, see below),
and thus it should behave almost identically to BKL. As it turns out,
it is even slightly better in some regimes.

\begin{figure}
\includegraphics[width=1\textwidth]{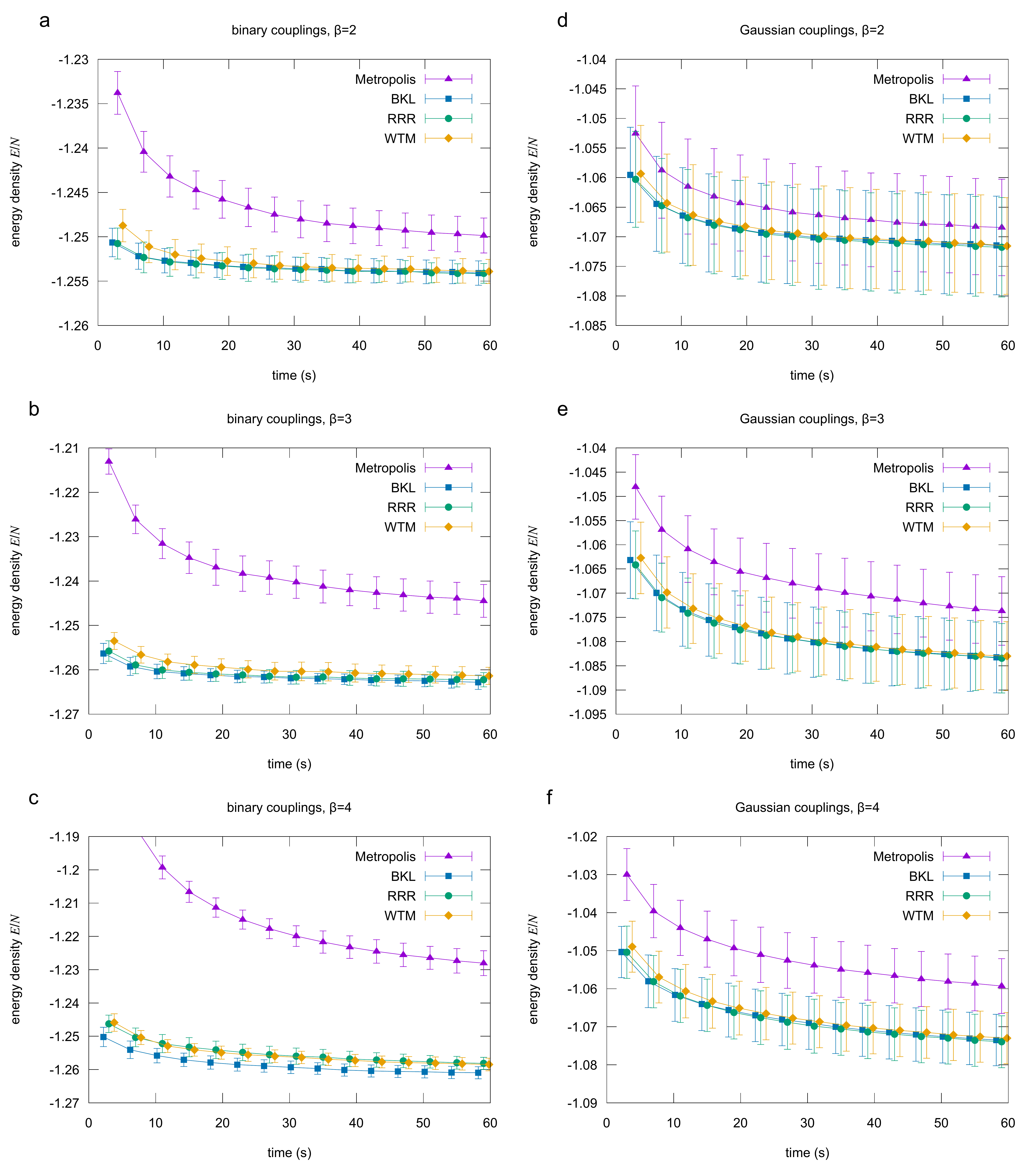}\caption{\label{fig:energy-vs-time-sparse}Energy density as a function of
time for four different algorithms at different values of $\beta$
on $20$ random regular graphs ($100$ runs for panel \textbf{a};
$2$ runs per algorithm per graph in all cases) with $N=10^{4}$ and
$K=3$, with either binary couplings (left column) or Gaussian couplings
(right column). See details in the text. The points and error bars
show the means and standard deviations (computed on bins of regular
size; they are slightly shifted relatively to each other for improved
readability). For the discrete case, the RRR algorithm performs best
at $\beta=2$ (this is visible in figures~\ref{fig:overlaps-and-accrates}c
and~\ref{fig:accpersec}a), RRR and BKL are about tied at $\beta=3$,
and BKL performs best at $\beta=4$ when the acceptance rate of RRR
drops (see also figures~\ref{fig:overlaps-and-accrates}e and~\ref{fig:accpersec}a);
WTM is slightly less optimized for this case and performs worse than
BKL. For the continuous case, BKL, RRR and WTM are all basically equivalent
at all temperatures (WTM is slightly worse), see also figure~\ref{fig:accpersec}b.}
\end{figure}

\begin{figure}
\includegraphics[width=1\textwidth]{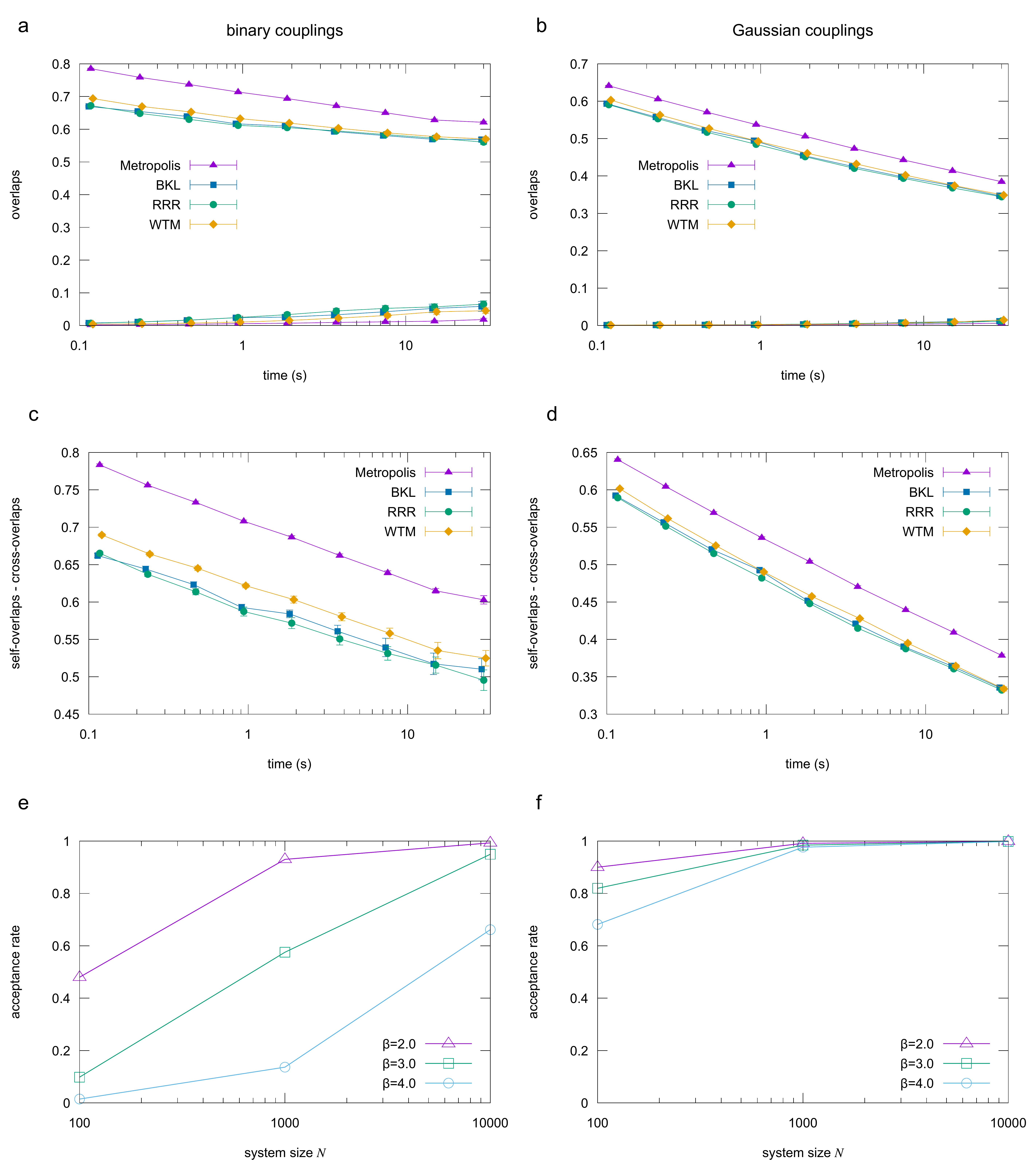}

\caption{\label{fig:overlaps-and-accrates}Overlaps vs time and RRR acceptance
rates for random regular graphs models at $K=3$ with either binary
couplings (left column) or Gaussian couplings (right column). Top
row: average self-overlaps (top curves) and cross-overlaps (bottom
curves) for $N=10^{4}$ and $\beta=2$, for four different algorithms
(same tests as panels \textbf{a} and \textbf{d} of figure~\ref{fig:energy-vs-time-sparse}).
Points are slightly shifted relative to each other for improved readability.
Middle row: average difference between self-overlaps and cross-overlaps,
same data as the two top panels. This shows that BKL is slightly better
than RRR in the discrete case at $\beta=2$, and that RRR, BKL and
WTM are equivalent in the continuous case; data for $\beta=3$ and
$\beta=4$ (not shown) agrees with the qualitative picture emerging
from figure~\ref{fig:energy-vs-time-sparse}. Bottom panels: acceptance
rate of RRR as a function of $N$ for different values of $\beta$
(the values at $N=10^{4}$ correspond to the plots in figure~\ref{fig:energy-vs-time-sparse}).
Error bars are smaller than the data points. The rates are higher
in the continuous case because the couplings can be small and some
spins can be flipped almost freely.}
\end{figure}

Figure~\ref{fig:energy-vs-time-sparse} shows the energy as a function
of time (for models with $N=10^{4}$, with $\beta$ ranging from $2$
to $4$), for $60$ seconds of simulations (thus nowhere near equilibrium).
Figure~\ref{fig:overlaps-and-accrates} shows the overlaps as a function
of time (for models with $N=10^{4}$ at $\beta=2$) and the RRR acceptance
rates at different values of $\beta$ and $N$.

Following \cite{bhatt_numerical_1988}, we computed both the self-overlaps
(the overlaps at different times, within a certain time interval,
for the same run) and the cross-overlaps (the overlaps for different
runs on the same graph at comparable times), and used their difference
as a measure of the distance from equilibrium. In both cases, we divided
the time in regular intervals in logarithmic scale (so that for example
in figure~\ref{fig:overlaps-and-accrates} the points displayed at
$t=15s$ used all samples collected between $15s$ and $30s$, while
the points displayed at $30s$ used the samples collected between
$30s$ and $60s$). The results make clear that the tests are indeed
very far from equilibrium, but they help discriminating better the
different algorithms, and are qualitatively consistent with the picture
that emerges from comparing the energies.

\begin{figure}

\includegraphics[width=1\textwidth]{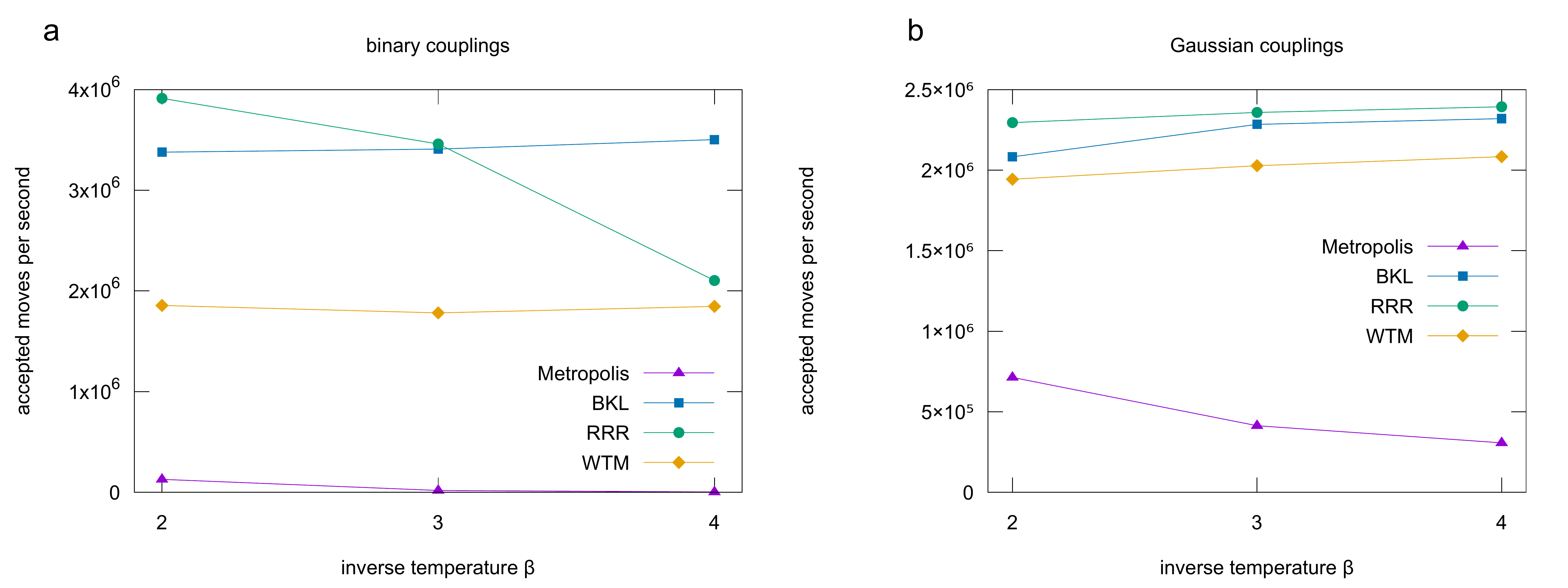}\caption{\label{fig:accpersec}Average moves per second for the data of figure~\ref{fig:energy-vs-time-sparse},
as a function of the inverse temperature $\beta$. Error bars are
smaller than the data points. RRR becomes worse than BKL only at low
temperatures for the binary couplings case.}

\end{figure}

Figure~\ref{fig:accpersec} shows the number of moves performed on
average per second for each algorithm, as a function of $\beta$,
again confirming the overall picture.

As mentioned above, since on these models BKL is essentially identical
to RRR in the very-high-acceptance-rate regime algorithmically, the
slight advantage of RRR over BKL in the discrete case at $\beta=2$
(which, although not clearly visible in figure~\ref{fig:energy-vs-time-sparse}a,
is visible in figure~\ref{fig:overlaps-and-accrates}c and even more
clearly in figure~\ref{fig:accpersec}a) can only be ascribed to
the computation of the number of moves to skip (step 1 in section~\ref{sec:BKL}).

Figures~\ref{fig:energy-vs-time-sparse} and~\ref{fig:accpersec}
show that only at $\beta=4$ in the discrete case RRR performs slightly
but measurably worse than BKL, since in that case the RRR acceptance
rate dropped to $0.66$ (figure~\ref{fig:overlaps-and-accrates}e).
For the continuous case, instead, even at these low temperatures the
acceptance rate is still greater than $0.99$ (figure~\ref{fig:overlaps-and-accrates}f),
and still better than BKL in terms of moves per second: this is due
to the fact that a fraction of the variables have nearly $0$ energy
shift cost and are thus selected with higher probability, while in
discrete RRG case no such cases exists since $K$ is odd. In general,
the bottom panels in figure~\ref{fig:overlaps-and-accrates} show
that, for any given $\beta$, the RRR acceptance rate increases monotonically
with $N$, as expected from the arguments given in section~\ref{sec:RRR}.

\subsection{Quantum Monte Carlo tests\label{subsec:Tests-Quantum}}

In the second set of experiments, we used the Suzuki-Trotter transformation
to study a quantum system of spins in a transverse magnetic field.
We will only sketch the method here, see e.g.~\cite{krzakala_path-integral_2008}
for a nice and more thorough introduction. Consider the following
Hamiltonian operator:
\begin{equation}
\hat{H}=-\sum_{ij}J_{ij}\hat{\sigma}_{i}^{z}\hat{\sigma}_{j}^{z}-\Gamma\sum_{i}\hat{\sigma}_{i}^{x}
\end{equation}
where $\hat{\sigma}_{i}^{z}$ is the spin operator (Pauli matrix)
in the longitudinal direction $z$, $\hat{\sigma}_{i}^{x}$ is the
spin operator in the transverse direction $x$, and $\Gamma\ge0$
is a magnetic field. The goal is to study the statistical mechanics
properties of the system at inverse temperature $\beta$, i.e.~the
partition function $Z=\mathrm{Tr}e^{-\beta\hat{H}}$ and the average
value of the observables $\left\langle \hat{O}\right\rangle =Z^{-1}\mathrm{Tr}\hat{O}e^{-\beta\hat{H}}$.
The well-known Suzuki-Trotter transformation~\cite{suzuki1976relationship}
allows to address this problem by using an effective classical Hamiltonian
of Ising spins, with an additional dimension with periodic boundary
conditions. The equivalence is realized when the size $M$ of this
additional (``imaginary'', or ``Trotter'') dimension diverges.
The effective classical Hamiltonian is written as:
\begin{equation}
H_{\mathrm{eff}}=-\frac{1}{M}\sum_{k}\sum_{ij}J_{ij}\sigma_{ik}\sigma_{jk}-\gamma\sum_{ki}\sigma_{ik}\sigma_{i\left(k+1\right)}\label{eq:H_eff}
\end{equation}
where $\sigma_{ik}\in\left\{ -1,1\right\} $ are classical spins,
with $i\in\left\{ 1,\dots,N\right\} $, $k\in\left\{ 1,\dots,M\right\} $
and the periodic condition $\sigma_{i\left(M+1\right)}\equiv\sigma_{i1}$,
and where $\gamma=\frac{1}{2\beta}\log\left(\coth\left(\frac{\beta\Gamma}{M}\right)\right)$.
Therefore, the transformation consists in replicating $M$ times the
longitudinal part of the original system (the part of the Hamiltonian
which commutes with the $\hat{\sigma}_{i}^{z}$), and adding pairwise
nearest-neighbor ferromagnetic interactions along the Trotter dimension
between the corresponding replicated spins. Hereafter, we refer to
the system ``slices'' at fixed $k$ as to the ``Trotter replicas''.

If such a system can be sampled efficiently, the average value of
the observables can be easily computed. In particular, the average
value of the energy density can be computed as:
\begin{equation}
\frac{\left\langle \hat{H}\right\rangle }{N}=\left\langle -\frac{1}{MN}\sum_{k}\sum_{ij}J_{ij}\sigma_{ik}\sigma_{jk}-\Gamma\left(\cosh\left(2\beta\gamma\right)-\left(\frac{1}{MN}\sum_{ki}\sigma_{ik}\sigma_{i\left(k+1\right)}\right)\sinh\left(2\beta\gamma\right)\right)\right\rangle .\label{eq:Hamiltonian-density}
\end{equation}

The interactions along the Trotter dimension all have connectivity
$2$, and all couplings are equal: therefore, that part of the Hamiltonian
$H_{\mathrm{eff}}$ describes a part of the model with small connectivity,
and we can thus apply the RRR method (and since the energy levels
are discrete, we can use the specialized discrete version for this
case). It is interesting to notice that, at small values of $\Gamma$,
the corresponding transverse interactions $\gamma$ diverge (as $\Gamma\to0$
the system becomes classical and the replicas collapse); therefore,
in that regime, these interactions dominate, and accounting for them
through a prior rather than the rejection rate seems particularly
promising. The BKL and WTM methods, on the other hand, would need
to take into account all the connections of each spin, which are extensive,
and are thus those methods are not efficient in this case (we verified
that they are indeed orders of magnitude slower than standard Metropolis
in the settings we tested).

This formalism is not only used to study quantum physical systems
(see e.g.~\cite{suzuki1977monte,nakamura2008spin}), but it is also
at the basis of some quantum annealing optimization techniques, in
which the main idea is that the system can be made to escape local
minima of the energy landscape via the tunneling effect by introducing
a transverse field in an otherwise classical problem, rather than
by thermal fluctuations. The general scheme is to simulate a system
at low temperature but with an initially strong transverse field $\Gamma$,
which is gradually lowered to $0$ in order to recover the original
classical system, hopefully in a low-energy configuration. The usage
of Monte Carlo sampling with the Suzuki-Trotter scheme in this context
is a well-known technique, see e.g.~\cite{santoro2002theory,martovnak2004quantum,battaglia2005optimization}.

\begin{figure}
\includegraphics[width=1\textwidth]{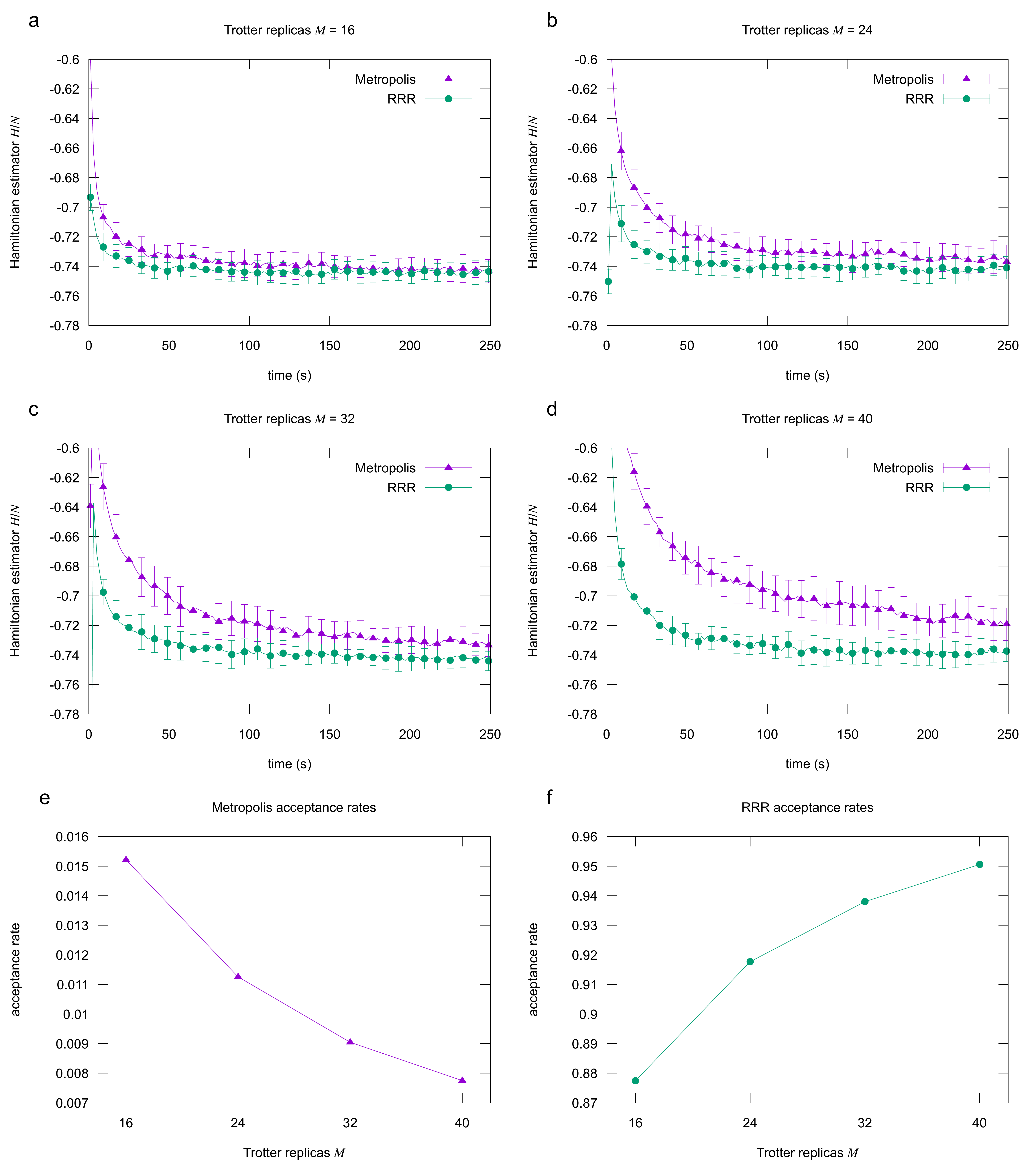}\caption{\label{fig:hamitonian-vs-time-quantum}Panels \textbf{a-d}: Instantaneous
Hamiltonian density estimation as a function of time for a fully connected
quantum spin model with random binary couplings and a transverse magnetic
field (see text for details), using different values of the Trotter
replica. The simulations were performed on SK models with $N=1024$
spins at fixed $\beta=2$ and $\Gamma=0.3$. The main plots compare
standard Metropolis and RRR, showing means and standard deviations
(after binning) on $20$ models, one run per each algorithm. Note
that the plotted quantity is the argument of the average in eq.~(\ref{eq:Hamiltonian-density}),
not the energy which governs the Monte Carlo process eq.~(\ref{eq:H_eff}),
which accounts for the non-monotonic behavior observed in the initial
iterations steps in some panels. RRR clearly converges to equilibrium
much faster. Panels \textbf{e-f}: the acceptance rate for the two
algorithms, as a function of the number of Trotter replicas $M$:
not only RRR has a much higher acceptance rate (even in terms of accepted
moves per second, RRR is at least a factor of $3$ larger), but the
behavior is opposite for the two algorithms.}

\end{figure}

We tested this approach on a Sherrington-Kirkpatrick (SK) fully-connected
model with random binary couplings $J_{ij}\in\left\{ -1/\sqrt{N},1/\sqrt{N}\right\} $,
$N=1024$ at fixed $\beta=2$ and $\Gamma=0.3$, measuring the value
of the ``instantaneous Hamiltonian density estimator'' (the term
between angle brackets in eq\@.~(\ref{eq:Hamiltonian-density}))
as a function of time. The results, shown in figure~\ref{fig:hamitonian-vs-time-quantum},
clearly indicate that the RRR method can equilibrate much faster than
standard Metropolis, and that the gain increases with larger values
of the Trotter dimension $M$. The acceptance rate increases with
$M$ for RRR because of the scaling factor $M^{-1}$ in the first
term of eq.~(\ref{eq:H_eff}), while the second term is accounted
for by the prior.

A more thorough exploration of the characteristics of the RRR method
on this kind of models, and a comparison with alternative Monte Carlo
algorithms specifically designed for this purpose like e.g.~\cite{nakamura2008efficient},
is reserved for a future work.

\subsection{Robust Ensemble tests\label{subsec:Tests-RE}}

In a final set of experiments, we applied the RRR method to the so-called
robust ensemble (RE), recently introduced in~\cite{baldassi2016_unreasonable}.
For a given Ising model with energy $E\left(\sigma\right)$, its RE
Hamiltonian at a given inverse temperature $\beta$ is defined on
a set of $M$ interacting replicas of the original system:
\begin{eqnarray}
H\left(\left\{ \sigma^{a}\right\} _{a=1}^{M};\beta,\gamma\right) & = & \sum_{a=1}^{M}E\left(\sigma^{a}\right)-\frac{1}{\beta}\sum_{i=1}^{N}\log\left(2\cosh\frac{\gamma}{2}\sum_{a=1}^{M}\sigma_{i}^{a}\right)\label{eq:H_RE}
\end{eqnarray}
where $\gamma$ is a control parameter. This Hamiltonian introduces
a measure which, compared to the original Gibbs distribution on $E\left(\sigma\right)$,
enhances the statistical weight of regions of the configuration space
with a large free entropy (i.e., roughly speaking, extensive regions
in which an exponential number of configurations have low energy),
with the parameter $\gamma$ indirectly controlling the scale of the
region (larger $\gamma$ values bring the focus to narrower regions).
One typical use of this measure is to use it within a ``scoping''
procedure, in which $\gamma$ is gradually increased. Therefore, the
large-$\gamma$ regime is of particular interest for this problem.
In terms of observables, one of the main quantities of interest is
the mean energy density across replicas, defined as:
\begin{equation}
\frac{\bar{E}}{N}=\left\langle \frac{1}{M}\sum_{a=1}^{M}E\left(\sigma^{a}\right)\right\rangle .\label{eq:mean-replica-energy-RE}
\end{equation}

This system is formally similar to the Quantum Monte Carlo of the
previous section,\footnote{This remarkable fact will be the subject of a future separate work.}
and we can use the RRR method to almost entirely account for the effect
of the interaction between replicas (the second term in eq.~(\ref{eq:H_RE})).
Note however that for this model the topology of this interaction
is different (it's fully-connected rather than a loop), and methods
based on flipping entire clusters of spins along a replicated dimensions
such as~\cite{nakamura2008efficient} are not applicable.

\begin{figure}

\includegraphics[width=1\textwidth]{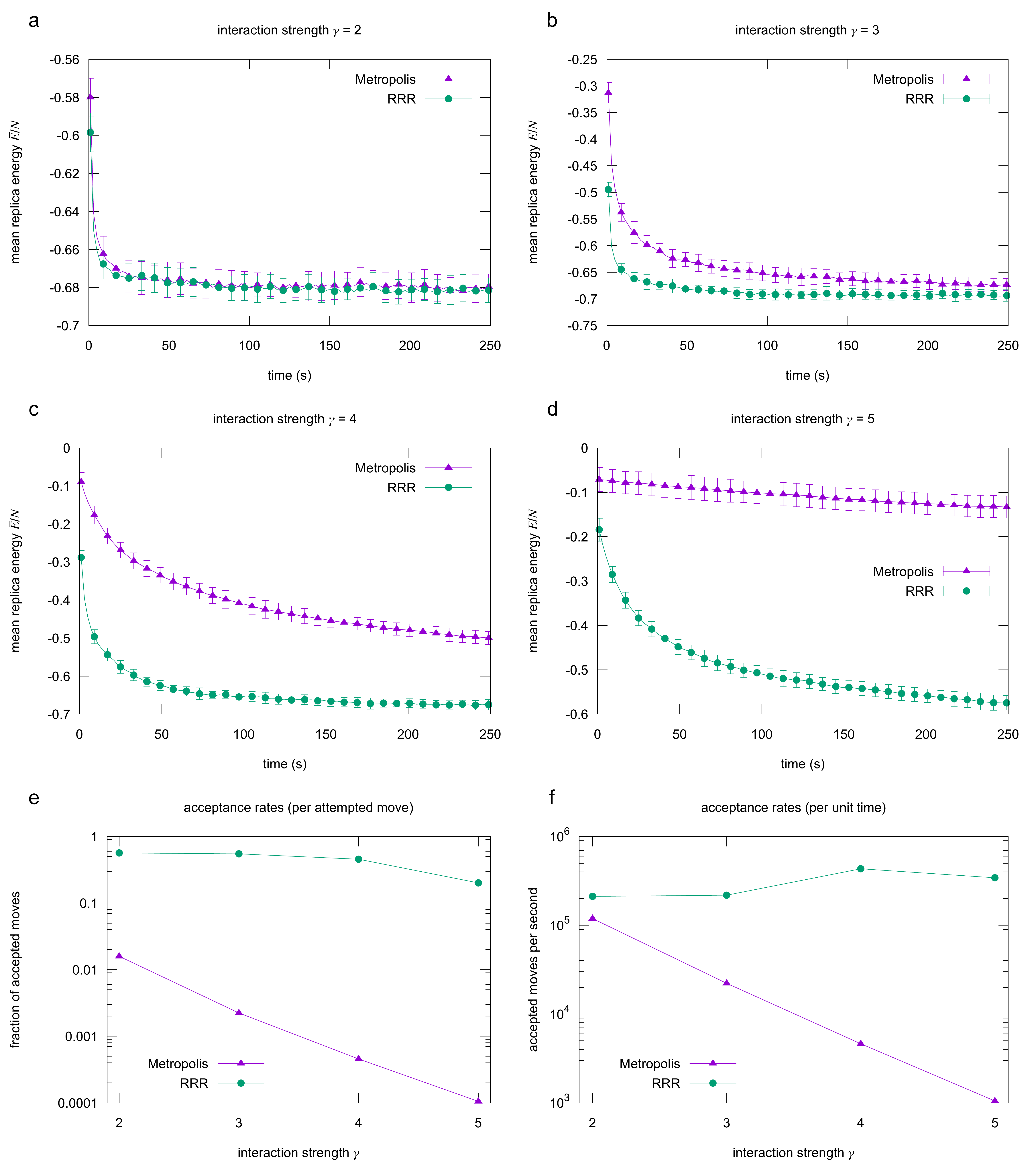}\caption{\label{fig:energy-vs-time-RE}Panels \textbf{a-d}: Mean energy density
across replicas (eq.~(\ref{eq:mean-replica-energy-RE})) as a function
of time with robust ensemble distribution (see text for details),
using different values of the interaction parameter $\gamma$. The
simulations were performed on SK models with $N=1024$ spins at fixed
$\beta=0.4$ and $M=5$. The main plots compare standard Metropolis
and RRR, showing means and standard deviations (after binning) on
$20$ models, one run per each algorithm. RRR clearly converges to
equilibrium much faster at large $\gamma$. Panel \textbf{e}: the
acceptance rates for the two algorithms, as a function of the interaction
parameter $\gamma$, in logarithmic scale: RRR is much less affected
by $\gamma$ since it absorbs its effect in the prior. Panel \textbf{f}:
the number of accepted moves per second for the two algorithms, in
logarithmic scale.}

\end{figure}

As in the previous section, we tested this approach on a replicated
Sherrington-Kirkpatrick fully-connected model with random binary couplings
$J_{ij}\in\left\{ -1/\sqrt{N},1/\sqrt{N}\right\} $, using $N=1024$,
$M=5$ replicas at fixed $\beta=0.4$ and varying $\gamma$. The results
are shown in figure~\ref{fig:energy-vs-time-RE}. As expected, RRR
is able to absorb most of the effect of the interaction in the prior,
and therefore its acceptance rate is almost constant up to very large
values of $\gamma$, while that of standard Metropolis drops dramatically.
This is true both in terms of accepted moves per attempted spin flip
(figure~\ref{fig:energy-vs-time-RE}e) and of accepted moves per
second (figure~\ref{fig:energy-vs-time-RE}f). As a result, RRR is
clearly advantageous in this large-$\gamma$ regime.

\section{Conclusions}

We have presented a Monte Carlo Markov Chain method, called reduced-rejection-rate
Monte Carlo (RRR), which extends the realm of applicability of rejection-free
methods: by transforming a kinetic Monte Carlo approach into a choice
for a prior, we were able to show that it is possible to improve over
the performance of a naïve Metropolis scheme by reducing the rejection
rate on a class of models. While rejection-free methods such as BKL
and WTM are effective (at low temperatures) for models with low connectivity,
the RRR method can also be applied to models in which only a component
of the Hamiltonian has that characteristic. For such models, the RRR
method can, in many cases, almost remove the rejection rate associated
with that component.

We demonstrated this by numerical experiments on Ising spin models,
first by showing that RRR indeed nearly eliminates the rejection rate
on sparse models (and is thus almost equivalent to BKL) in a wide
range of regimes; then by applying it to two models (quantum models
and robust-ensemble models) in which only a part of the Hamiltonian
is sparse, and showing that the reduction of the rejection rates leads
to improved dynamics with respect to a naïve Metropolis scheme when
the sparse part dominates. The experiments were mostly exploratory
and aimed at demonstrating the feasibility of the method, and are
by no means exhaustive: further work is needed (and planned) to employ
this method in practically relevant applications and as a component
within more general algorithmic schemes (e.g. simulated annealing
or parallel tempering). Since the code is generic and extensible,
and publicly available, it will also easily allow for a more wide-range
exploration of the effectiveness of the technique to other type of
sampling and optimization problems.

The method itself is not restricted to Ising spin models: just like
the rejection-free methods it is derived from, it could be straightforwardly
applied to models with multiple states per variable (Potts-like models).
Of course, this would come at an additional algorithmic cost. In general,
the results presented here suggest that such scheme would certainly
be convenient whenever the problem is dominated by a sparse component
for which the kinetic Monte Carlo approach is better than standard
Metropolis, but is also subject to additional dense interactions.
\begin{acknowledgments}
I am deeply indebted to G.~Parisi for some crucial observations on
an early draft of this manuscript, to R.~Zecchina for suggestions
and discussions, and to the anonymous referees that reviewed this
work for their precious comments. I also wish to thank M.~Zamparo
and A.~Pagnani for helpful suggestions and discussions.
\end{acknowledgments}


\begin{thebibliography}{10}

\bibitem{bortz_new_1975}
A.B. Bortz, M.H. Kalos, and J.L. Lebowitz.
\newblock A new algorithm for {Monte} {Carlo} simulation of {Ising} spin
  systems.
\newblock {\em Journal of Computational Physics}, 17(1):10--18, January 1975.
\newblock \href {http://dx.doi.org/10.1016/0021-9991(75)90060-1}
  {\path{doi:10.1016/0021-9991(75)90060-1}}.

\bibitem{baldassi2016_unreasonable}
Carlo Baldassi, Christian Borgs, Jennifer~T. Chayes, Alessandro Ingrosso, Carlo
  Lucibello, Luca Saglietti, and Riccardo Zecchina.
\newblock Unreasonable effectiveness of learning neural networks: From
  accessible states and robust ensembles to basic algorithmic schemes.
\newblock {\em Proceedings of the National Academy of Sciences}, 2016.
\newblock \href {http://dx.doi.org/10.1073/pnas.1608103113}
  {\path{doi:10.1073/pnas.1608103113}}.

\bibitem{RRR-code}
{RRR Monte Carlo code}.
\newblock \url{https://github.com/carlobaldassi/RRRMC.jl}.

\bibitem{dall_faster_2001}
Jesper Dall and Paolo Sibani.
\newblock Faster {Monte} {Carlo} simulations at low temperatures. {The} waiting
  time method.
\newblock {\em Computer Physics Communications}, 141(2):260--267, November
  2001.
\newblock \href {http://dx.doi.org/10.1016/S0010-4655(01)00412-X}
  {\path{doi:10.1016/S0010-4655(01)00412-X}}.

\bibitem{krauth_statistical_2006}
Werner Krauth.
\newblock {\em Statistical mechanics: algorithms and computations}.
\newblock Number~13 in Oxford master series in physics {Statistical},
  computational, and theoretical physics. Oxford Univ. Press, Oxford, 2006.
\newblock OCLC: 254061236.

\bibitem{kirkpatrick1983optimization}
S.~Kirkpatrick, C.~D. Gelatt, and M.~P. Vecchi.
\newblock Optimization by {Simulated} {Annealing}.
\newblock {\em Science}, 220(4598):671--680, May 1983.
\newblock \href {http://dx.doi.org/10.1126/science.220.4598.671}
  {\path{doi:10.1126/science.220.4598.671}}.

\bibitem{santoro2002theory}
G.~E. Santoro.
\newblock Theory of {Quantum} {Annealing} of an {Ising} {Spin} {Glass}.
\newblock {\em Science}, 295(5564):2427--2430, March 2002.
\newblock \href {http://dx.doi.org/10.1126/science.1068774}
  {\path{doi:10.1126/science.1068774}}.

\bibitem{voter2007introduction}
Arthur~F. Voter.
\newblock {INTRODUCTION} {TO} {THE} {KINETIC} {MONTE} {CARLO} {METHOD}.
\newblock In Kurt~E. Sickafus, Eugene~A. Kotomin, and Blas~P. Uberuaga,
  editors, {\em Radiation {Effects} in {Solids}}, volume 235, pages 1--23.
  Springer Netherlands, Dordrecht, 2007.
\newblock \href {http://dx.doi.org/10.1007/978-1-4020-5295-8_1}
  {\path{doi:10.1007/978-1-4020-5295-8_1}}.

\bibitem{wang1990cluster}
Jian-Sheng Wang and Robert~H. Swendsen.
\newblock Cluster {Monte} {Carlo} algorithms.
\newblock {\em Physica A: Statistical Mechanics and its Applications},
  167(3):565--579, September 1990.
\newblock \href {http://dx.doi.org/10.1016/0378-4371(90)90275-W}
  {\path{doi:10.1016/0378-4371(90)90275-W}}.

\bibitem{wolff1989collective}
Ulli Wolff.
\newblock Collective {Monte} {Carlo} {Updating} for {Spin} {Systems}.
\newblock {\em Physical Review Letters}, 62(4):361--364, January 1989.
\newblock \href {http://dx.doi.org/10.1103/PhysRevLett.62.361}
  {\path{doi:10.1103/PhysRevLett.62.361}}.

\bibitem{earl2005parallel}
David~J. Earl and Michael~W. Deem.
\newblock Parallel tempering: {Theory}, applications, and new perspectives.
\newblock {\em Physical Chemistry Chemical Physics}, 7(23):3910, 2005.
\newblock \href {http://dx.doi.org/10.1039/b509983h}
  {\path{doi:10.1039/b509983h}}.

\bibitem{landau_guide_2009}
David~P. Landau and K.~Binder.
\newblock {\em A guide to {Monte} {Carlo} simulations in statistical physics}.
\newblock Cambridge University Press, Cambridge; New York, 3rd ed edition,
  2009.
\newblock OCLC: ocn444371657.

\bibitem{suzuki1976relationship}
M.~Suzuki.
\newblock Relationship between d-{Dimensional} {Quantal} {Spin} {Systems} and
  (d+1)-{Dimensional} {Ising} {Systems}: {Equivalence}, {Critical} {Exponents}
  and {Systematic} {Approximants} of the {Partition} {Function} and {Spin}
  {Correlations}.
\newblock {\em Progress of Theoretical Physics}, 56(5):1454--1469, November
  1976.
\newblock \href {http://dx.doi.org/10.1143/PTP.56.1454}
  {\path{doi:10.1143/PTP.56.1454}}.

\bibitem{wong_efficient_1980}
C.~K. Wong and M.~C. Easton.
\newblock An {Efficient} {Method} for {Weighted} {Sampling} without
  {Replacement}.
\newblock {\em SIAM Journal on Computing}, 9(1):111--113, February 1980.
\newblock \href {http://dx.doi.org/10.1137/0209009}
  {\path{doi:10.1137/0209009}}.

\bibitem{matias_dynamic_2003}
Yossi Matias, Jeffrey~Scott Vitter, and Wen-Chun Ni.
\newblock Dynamic {Generation} of {Discrete} {Random} {Variates}.
\newblock {\em Theory of Computing Systems}, 36(4):329--358, August 2003.
\newblock \href {http://dx.doi.org/10.1007/s00224-003-1078-6}
  {\path{doi:10.1007/s00224-003-1078-6}}.

\bibitem{bhatt_numerical_1988}
R.~N. Bhatt and A.~P. Young.
\newblock Numerical studies of {Ising} spin glasses in two, three, and four
  dimensions.
\newblock {\em Physical Review B}, 37(10):5606--5614, April 1988.
\newblock \href {http://dx.doi.org/10.1103/PhysRevB.37.5606}
  {\path{doi:10.1103/PhysRevB.37.5606}}.

\bibitem{krzakala_path-integral_2008}
Florent Krzakala, Alberto Rosso, Guilhem Semerjian, and Francesco Zamponi.
\newblock Path-integral representation for quantum spin models: {Application}
  to the quantum cavity method and {Monte} {Carlo} simulations.
\newblock {\em Physical Review B}, 78(13), October 2008.
\newblock \href {http://dx.doi.org/10.1103/PhysRevB.78.134428}
  {\path{doi:10.1103/PhysRevB.78.134428}}.

\bibitem{suzuki1977monte}
M.~Suzuki, S.~Miyashita, and A.~Kuroda.
\newblock Monte {Carlo} {Simulation} of {Quantum} {Spin} {Systems}. {I}.
\newblock {\em Progress of Theoretical Physics}, 58(5):1377--1387, November
  1977.
\newblock \href {http://dx.doi.org/10.1143/PTP.58.1377}
  {\path{doi:10.1143/PTP.58.1377}}.

\bibitem{nakamura2008spin}
Tota Nakamura and Yoichi Nishiwaki.
\newblock Spin-lattice model of magnetoelectric transitions in {RbCoBr} 3.
\newblock {\em Physical Review B}, 78(10), September 2008.
\newblock \href {http://dx.doi.org/10.1103/PhysRevB.78.104422}
  {\path{doi:10.1103/PhysRevB.78.104422}}.

\bibitem{martovnak2004quantum}
Roman Marto{\v{n}}{\'a}k, Giuseppe~E. Santoro, and Erio Tosatti.
\newblock Quantum annealing of the traveling-salesman problem.
\newblock {\em Physical Review E}, 70(5), November 2004.
\newblock \href {http://dx.doi.org/10.1103/PhysRevE.70.057701}
  {\path{doi:10.1103/PhysRevE.70.057701}}.

\bibitem{battaglia2005optimization}
Demian~A. Battaglia, Giuseppe~E. Santoro, and Erio Tosatti.
\newblock Optimization by quantum annealing: Lessons from hard satisfiability
  problems.
\newblock {\em Phys. Rev. E}, 71:066707, Jun 2005.
\newblock \href {http://dx.doi.org/10.1103/PhysRevE.71.066707}
  {\path{doi:10.1103/PhysRevE.71.066707}}.

\bibitem{nakamura2008efficient}
Tota Nakamura.
\newblock Efficient {Monte} {Carlo} {Algorithm} in {Quasi}-{One}-{Dimensional}
  {Ising} {Spin} {Systems}.
\newblock {\em Physical Review Letters}, 101(21), November 2008.
\newblock \href {http://dx.doi.org/10.1103/PhysRevLett.101.210602}
  {\path{doi:10.1103/PhysRevLett.101.210602}}.

\end{thebibliography}
\end{document}